# A Physiological Sensor-Based Android Application Synchronized with a Driving Simulator for Driver Monitoring

David González-Ortega *, Francisco Javier Díaz-Pernas, Mario Martínez-Zarzuela and Míriam Antón-Rodríguez

Department of Signal Theory, Communications and Telematics Engineering, Telecommunications Engineering School, University of Valladolid, 47011 Valladolid, Spain; pacper@tel.uva.es (F.J.D.-P.); marmar@tel.uva.es (M.M.-Z.); mirant@tel.uva.es (M.A.-R.)
* Correspondence: davgon@tel.uva.es; Tel.: +34-983-423-000 (ext. 5552)

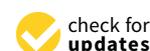

**Abstract:** In this paper, we present an Android application to control and monitor the physiological sensors from the Shimmer platform and its synchronized working with a driving simulator. The Android app can monitor drivers and their parameters can be used to analyze the relation between their physiological states and driving performance. The app can configure, select, receive, process, represent graphically, and store the signals from electrocardiogram (ECG), electromyogram (EMG) and galvanic skin response (GSR) modules and accelerometers, a magnetometer and a gyroscope. The Android app is synchronized in two steps with a driving simulator that we previously developed using the Unity game engine to analyze driving security and efficiency. The Android app was tested with different sensors working simultaneously at various sampling rates and in different Android devices. We also tested the synchronized working of the driving simulator and the Android app with 25 people and analyzed the relation between data from the ECG, EMG, GSR, and gyroscope sensors and from the simulator. Among others, some significant correlations between a gyroscope-based feature calculated by the Android app and vehicle data and particular traffic offences were found. The Android app can be applied with minor adaptations to other different users such as patients with chronic diseases or athletes.

## 1. Introduction

Mobile devices are increasingly present in our daily life. With the widespread use of these devices, there is a huge demand for applications to address the need of millions of users in many fields such as education, health, business, commerce, and entertainment. Android is the clear leader in the global smartphone market as it is being used on 85% of smartphones all over the world [1]. Besides, Android users are very receptive to new applications thus this platform stands out as the best option to develop a mobile application.

Wearable sensors have gained a lot of importance in different fields such as education [2] and health [3,4]. These sensors play a main part in ambient intelligence research on its way to revolutionize daily human life by making people's environments sensitive, flexible, and responsive [5]. Regarding health, physiological sensors can be used to assess the human functional state [6]. Sensors should have the features required for m-health (mobile health) to address its four main challenges: continuous monitoring, full involvement, interoperability, and disappearing interaction [7]. They have to be lightweight and energy efficient [8]. Safety requirements are also important [8,9]. Sensors have to obtain accurate and reliable data, as erroneous measurements may lead to a misinterpretation of the data [10]. Keeping the costs low is a key aspect to allow sensors to be widely adopted. Most current physiological sensors use ZigBee or Bluetooth protocols to transmit wireless packets [11]. The advantages of these two protocols are small power requirements, low electromagnetic wave strength, and low cost [12]. Bluetooth is delivered in the 2.4 GHz band with a transmission speed of up to 25 Mbps (Bluetooth 4.0) and is not subject to electromagnetic interference as it uses low-power transmission. As a result, it is less harmful to the human body.

Mobile health applications interacting with physiological sensors can monitor electrocardiogram (ECG), electromyogram (EMG), respiration rate, body temperature, and limb and body motion of a person for a long term, without the need to be hospitalized, for prevention, early detection, and monitoring of cardiovascular, neuro-degenerative, and other chronic diseases, elderly assistance at home, fitness and wellness, motor rehabilitation assistance, physical activity, and gestures detection [11,13–17].

Physiological sensors from the Shimmer (an acronym for Sensing Health with Intelligence, Modularity, Mobility and Experimental Reusability) platform (Shimmer, Dublin, Ireland) include ECG, EMG, galvanic skin response (GSR), as well as accelerometers, a magnetometer, and a gyroscope [18]. They are small and lightweight devices that can obtain accurate data for biomedical research. Researchers can take advantage of several features of this sensing platform, such as its integrated peripherals, open software, modular expansion, specific power management hardware, and a library of applications supported with platform validation. These features make Shimmer stand out over many other medical platforms [19]. Burns et al. [20] validated the physiological signals (ECG, EMG, and GSR) of the Shimmer platform against known commercial systems. O'Donovan et al. [21] studied the utility of the Shimmer platform for temporal gait analysis. They concluded that Shimmer is an ideal solution for in-home and ambulatory monitoring of temporal gait parameters. Abate et al. [22] presented a smartphone-based system that monitors the motions of patients, recognizes a fall, and automatically sends a request for help to the caregivers. A Shimmer sensing unit was used for the acquisition of the motion data. Richer at al. [23] developed an Android application that calculates the heart rate and cadence of a cyclist using the ECG and EMG sensors of the Shimmer platform. The application can be used to monitor and analyze the cyclist's training sessions.

A sensing system that can monitor a driver's level of attention and alert him or her if this level is below a threshold could prevent some road accidents. The correlation between driver physiological signals and drowsiness has been studied by many researchers [24–26]. Steering wheel behavior has also been used to determine driver fatigue. The recognition of the way a vehicle is steered and, if necessary, its notification to drivers in real time would be crucial to achieve safer driving since fatal accidents are often caused by dangerous vehicle maneuvers, such as rapid turns and fast lane changes, or by a drowsy period in which the driver does not turn the steering wheel when he or she is about to fall asleep. Li et al. [27] proposed a system to monitor the level of driver fatigue under real driving conditions. It is based on the data of steering wheel angles collected from sensors mounted on the steering lever. Liu et al. [28] presented a system to recognize dangerous vehicle steering based on the gyroscope embedded in a smartphone. Three steering maneuvers, including turns, lane-changes, and U-turns, were defined, and a Fast Dynamic Time Warping (FastDTW) algorithm was adopted to recognize the vehicle steering.

In the past few years, driving simulators have been used in many research studies. With a simulator, it is possible to assess the driver behavior in situations that cannot be replicated on real roads [29]. The suitability of simulators to assess the driver behavior and as a tool to learn efficient and safe driving skills has been verified. Sullman et al. [30] showed the potential of training in efficient driving techniques using a simulator for bus drivers. Jamson et al. [31] analyzed the tradeoff between driving efficiency and safety through the use of in-vehicle embedded systems that advise on the use of the accelerator pedal in a simulator. They proved that these systems could improve driving efficiency. A driving simulator can include scenarios to analyze the level of driving efficiency and safety, the relation between both and the results obtained as a function of them. Besides, driving simulators enable the study of how certain factors can influence driving abilities, such as fatigue, physical and psychological state of the driver, age, and consumption of medicines, drugs, or alcohol. Driving simulators can easily store vehicle-based measures such as speed, rpm (revolutions per minute), or fuel consumption. The integration of sensors in a simulator allows to obtain physiological measures. Awais et al. [32] proposed an accurate driver drowsiness system combining EEG and ECG sensors, which was tested in a driving simulator. Other than these two types of measures, researchers have also used subjective measures where drivers are asked to rate their level of drowsiness either verbally or through a questionnaire [33]. All the measures can be analyzed together to see how they correlate and their influence on the number of driving errors and traffic offences. Samiee et al. [34] presented a drowsiness detection approach that combines vehicle-based measures (lateral position and steering wheel angle cues) and a physiological measure such as driver's eye status. They achieved a reliable system robust to the input signal loss. Krajewski et al. [35] studied the influence of fatigue in steering wheel behavior in a driving simulator. Wang et al. [36] analyzed nineteen driving behavior variables and four eye feature variables, together with the subjective measure of fatigue given by a questionnaire, to determine fatigue using a multilevel ordered logit model, an ordered logit model, and an artificial neural network. Yang et al. [32] proposed a driver fatigue recognition model based on a dynamic Bayesian network, information fusion, and multiple contextual and physiological features. Their experiments highlighted the significance of physiological features to infer driver fatigue.

In this paper, we present an Android health application to control the Shimmer sensors through a Bluetooth connection, by monitoring their signals simultaneously and processing and storing them in different files as a function of the sensors and the monitored user. These files can be sent by e-mail or shared with other Android applications such as Dropbox or Google Drive. This application was briefly introduced in [37]. The Android app also processes the angular speed of a gyroscope, which can be placed on the steering wheel of a vehicle, and obtains many features of interest to analyze driving performance, which was added from the application presented in [37]. Besides, we synchronized the Android app with a driving simulator that we developed using the Unity game engine and C# language. The simulator includes different urban and interurban scenarios and stores information about the committed traffic offences. A Unity app had to be developed and installed in the Android device to bridge between the Android app for the Shimmer sensors and the PC driving simulator. The Android app was tested with different sensors working simultaneously at various sampling rates in different Android devices. The synchronized working between the driving simulator and the Android app was tested with 19 people driving in four different scenarios using automatic and manual gear shift. We analyzed the relation between data from the gyroscope and the simulator, characteristics of the drivers and the subjective measure of their sleepiness using three questionnaires described in the literature.

The rest of the paper is organized as follows: Section 2 first introduces the physiological sensors from the Shimmer platform and the developed Android application. Later, it explains the gyroscope-based processing in the Android app to extract features of interest to analyze driving performance and the synchronization between the Android app and our driving simulator. Afterwards, Section 3 analyzes the experimental results, beginning with the synchronized working of the sensors in the Android app and following with the experiments carried out with the synchronized working of the driving simulator and the Android app and the analysis of the extracted data from both applications. Finally, Section 4 draws the main conclusions about the presented work.

## 2. Materials and Methods

### 2.1. Physiological Sensors from the Shimmer Platform

The Shimmer platform [18] provides a set of physiological sensors that can measure different human body parameters suitable for biomedical research. The signals are transmitted through a Bluetooth connection to a PC or a mobile device using an appropriate application. The Shimmer sensors are small, compact and weigh around 20 g. These features make them suitable for placement on the human body regions required to obtain the necessary measurements. All the sensors are controlled by an MSP430 microprocessor (Texas Instruments, Dallas, TX, USA) with a 16-bit Reduced Instruction Set Computer (RISC) architecture, which has low power consumption, especially during inactivity periods. The mainboard of the sensors has a 3-axis accelerometer with a range of $\pm 2$ g, $\pm 4$ g, $\pm 8$ g, or $\pm 16$ g. They have two radio modules based on the standards 802.15 (Bluetooth with Mitsumi WML-C46N CSR design and 802.15.4). Both modules can work simultaneously and stay disconnected in inactivity periods to reduce the power consumption. The CC2420 transceiver (Texas Instruments, Dallas, TX, USA) is used in the 802.15.4 module. Apart from the signal of the accelerometer, each sensor includes another module in its basic unit to extract physiological parameters such as ECG, EMG, GSR, or nine Degrees of Freedom (9DOF).

The ECG module can obtain the electrical activity of the heart. To that aim, four electrodes connected to the ECG module have to be placed strategically around the rib cage. Thus, the electrical activity of the heart beats is registered. ECG is a main tool in cardiology, very important for the diagnosis and monitoring of cardiovascular diseases. Figure 1a shows the Shimmer ECG module and Figure 1b shows the positions where the four electrodes (RA, LA, LL, and RL) have to be placed.

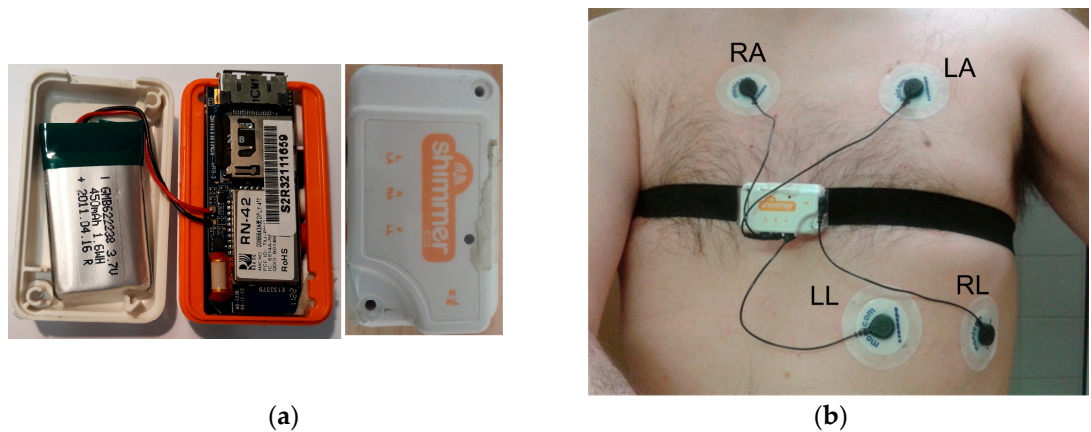

(**a**)　　　　　　　　　　　　　　　　　(**b**)

**Figure 1.** Shimmer electrocardiogram (ECG) sensor: (**a**) Shimmer ECG module; (**b**) Positions of the electrodes of the ECG sensor.

The EMG module can obtain the electrical activity related to the muscular contractions. It can be used to evaluate nerve conduction, the muscular response in an injured tissue, the activation level, and the biomechanics of human motion. Its electrodes have to be carefully placed to obtain correct values. The positive and negative electrodes have to be placed in parallel to the muscle fiber. The reference electrode has to be placed in an electrically neutral region of the body as far as possible from the monitored muscle. Figure 2 shows positions where the three electrodes can be placed. The reference electrode is placed on the wrist in the left image of Figure 2 and the positive and negative electrodes are placed close to one another on the biceps in the right image of Figure 2.

The GSR module can obtain the electrical signal caused by the changes in the skin resistance. The emotional state of a person determines such changes. The two electrodes have to be placed in two fingers of one hand as shown in Figure 3.

The 9DOF module combines a gyroscope with a magnetometer and is a powerful solution to kinematic detection. This module, together with the accelerometer incorporated in the mainboard of the sensors, can provide a static and dynamic orientation of inertial measurement and thus becomes a complex motion sensor. The gyroscope obtains the speed in the changes of the orientation, measured in degrees per second, with respect to the three axes. The accelerometer, magnetometer, and gyroscope have to be calibrated to obtain accurate data.

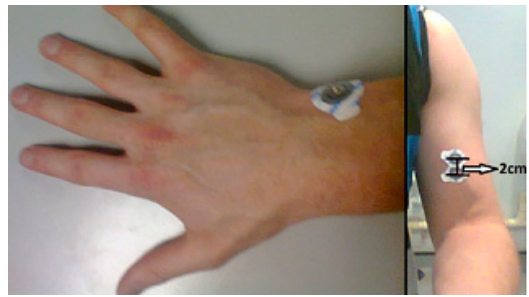

**Figure 2.** Position of the electrodes of the Shimmer electromyogram (EMG) sensor on the wrist (reference electrode) and the biceps.

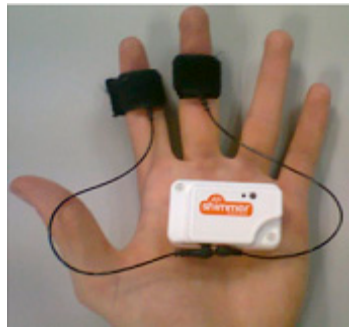

**Figure 3.** Position of the electrodes of the Shimmer galvanic skin response (GSR) sensor.

*2.2. Android Application for the Shimmer Sensors*

To develop the application, the Shimmer Android Instrument Driver Library (version 2.1) was used. Furthermore, a set of classes have been developed to interact with the Shimmer sensors, with the Bluetooth functionality, and with the storage memory to save the data in the corresponding files. The Shimmer library allows the communication with the Shimmer sensors through the Serial Port Profile (SPP) protocol. Android has used this protocol for Bluetooth communication since version 2.0 (API 5).

The Shimmer class in the Shimmer library provides, among other things, the interaction with the sensors, allowing to set up the sampling rate, change the range of the accelerometer, enable or disable some modules of the sensors, and connect several sensors simultaneously to an Android device using several instances of the class. This class is based on the Bluetooth interface provided by Android to set the connection with the particular sensor through the SPP. This protocol is in charge of emulating a connection through serial cable with the wireless Bluetooth technology. Each sensor connected with the application is represented through a Shimmer object (instance of the Shimmer class).

The application is structured in three activities: *MainActivity*, *ManageFiles*, and *ManageDevices*. The first activity deals with the Bluetooth settings. It allows to check the current state of the Bluetooth connection, change that state, search for nearby Bluetooth devices, and select them to establish a connection. The second activity manages the previously saved files. Those files have the data stored in past sessions organized as a function of the user. The files can be shared with other applications such as e-mail or cloud storage. The third activity is in charge of detecting the state of the Shimmer sensors, managing communications with the sensors, signal processing and representation, and data storage.

Depending on the sensor, the application can perform different processing tasks. With the ECG signals, the application obtains Heart Rate Variability (HRV) data. HRV is a measure of the beat-to-beat (R-R intervals) in the heart rate, which has been extensively used in the literature as its value depends on the physiological state of a person [24]. The EMG and GSR samples are grouped in sets of 20 samples. For each set, the mean and the standard deviation are calculated together with the total mean and standard deviation considering all the samples. The gyroscope signal of the 9DOF module is computed to obtain a large number of features of interest that will be detailed in the next subsection and analyzed in the experimental results.

The Graphical User Interface (GUI) of the application utilizes both the widget subclass (*View* class) and the layout subclass (*Viewgroup* class). The widgets are objects already developed that make programming easier and are directly included in the user interface, e.g., text fields and buttons. The layouts allow to contain the included widgets in an organized way, i.e., provide layout architectures, such as Linear, Tabular, and Relative. The GUI has four screens. Figure 4a shows the first screen (Bluetooth settings). It has different sections. The first section has the title and a menu button with only one option "Manage files", which allows to access the activity that is in charge of the files where the sensor data in the previous executions have been saved. The second section is a layout of the state of the Bluetooth that allows to change that state. The third section is a button that allows to search for nearby Bluetooth devices. The fourth section has a list of found Bluetooth devices. Finally, the fifth section "Manage Devices" has a button to access the next activity to manage the selected devices. Figure 4b shows the second screen (File management).

Figure 4c shows the third screen where the sensors are managed. Its first section has a menu button to select among the following options: "Start all" to set all the sensors in the transmission state, "Streaming" to begin the transmission of the sensors, "Disconnect all" to disconnect all the connected sensors, and "Go home" to return to the first activity. In particular, Figure 4c shows that all the four sensors are streaming. Figure 4d,e show the fourth screen where graphs can be seen with the online data obtained from the ECG and EMG sensors, respectively. As the sensors that obtain ECG and EMG, similarly to all the sensors, also includes an accelerometer, its graph is also shown in the screens. This graph includes three curves that correspond with the acceleration in the three axes. The acceleration in the x, y and z axes are represented in the red curve, green curve, and blue curve, respectively. The application needs some classes to work although they do not belong to any activity such as *GraphViews*, *DataLogging*, *Tools*, and *mySharedPrefs*. *Graph View* creates a view corresponding to a graph with the axes defined as a function of the signal represented in them. Its method *setData* receives an array with samples of a signal and goes across them representing each sample of the corresponding signal. *DataLogging* creates the files in the corresponding user folders, writes the header of the files and then the received data. The *Tools* activity provides the operations that can be applied to the sensors as a function of their state. *mySharedPrefs* stores the configuration of the sensors, recovers it and makes the same operations for each property of that configuration individually.

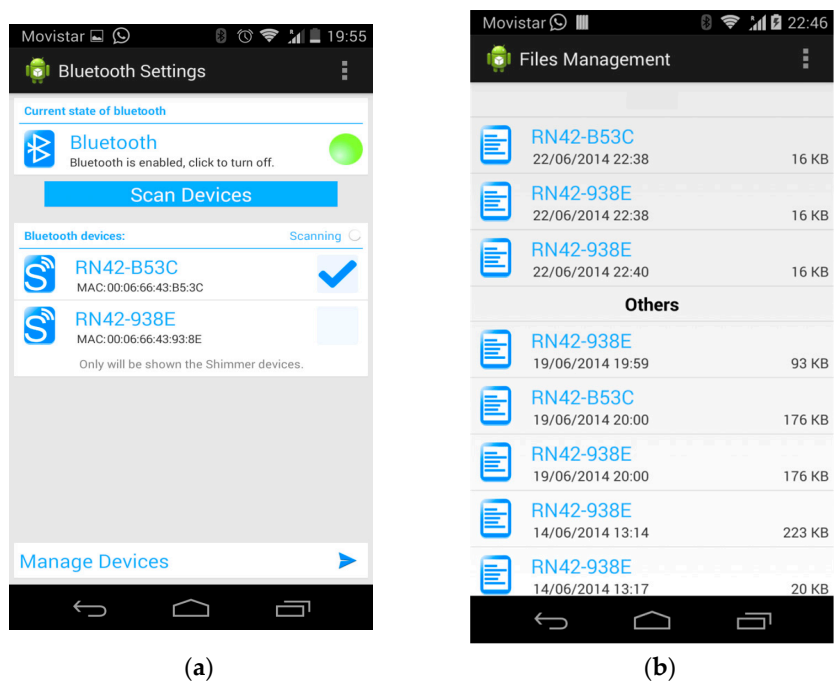

(**a**)　　(**b**)

**Figure 4.** *Cont.*

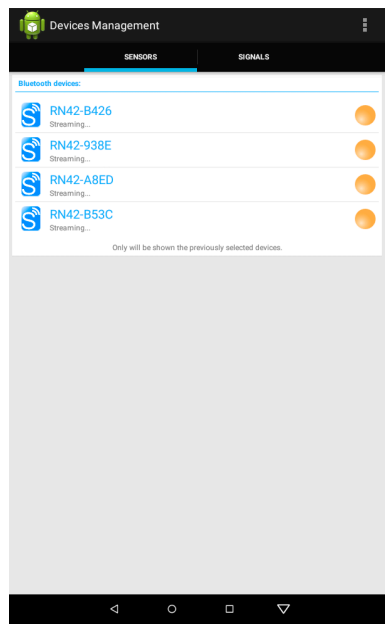

(**c**)

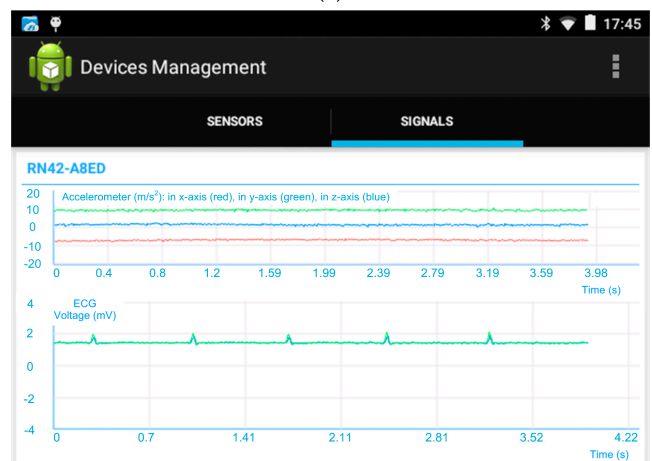

(**d**)

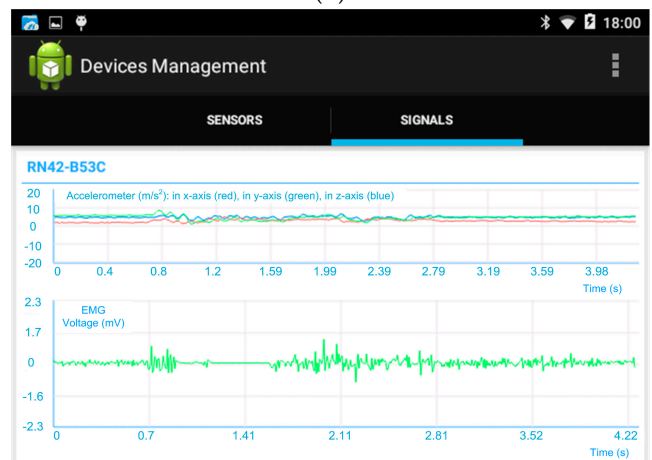

(**e**)

**Figure 4.** Five screens of the Graphical User Interface (GUI) of the application: (**a**) First screen of the GUI; (**b**) Second screen of the GUI; (**c**) Third screen of the GUI; (**d**) Fourth screen of the GUI with the accelerometer and ECG graphs; (**e**) Fourth screen of the GUI with the accelerometer and EMG graphs.

*2.3. Gyroscope-Based Processing in the Android App*

The Android app for the Shimmer sensors includes the computation and display of a large number of features of interest to analyze driving performance. These features were previously used in past research [34,36]. The 9DOF module is placed on the steering wheel aligned with its rotation axis. We aimed to measure the turning patterns of the steering wheel of the drivers as a function of their driving behavior using the gyroscope of the 9DOF module. The relevant angular speed of the gyroscope in such position will be in the z-axis. Its graph is shown in real time in the "Signals" tab of the app alongside with the graph of the data given by the accelerometer included in the 9DOF module, as shown in Figure 5.

Below these graphs, there are two buttons that have two styles: one for the on-state and another for the off-state. These styles are used to show in real time if a left or a right turn of the steering wheel is made. The current position of the steering wheel in degrees is calculated as follows:

$$Present\_position \text{ [degrees]} = previous\_position + present\_angular\_speed/F_s, \qquad (1)$$

where $F_s$ is the sampling rate selected for the gyroscope.

When the current position reaches absolute values greater than 360º, a variable that stores the number of complete turns is increased. With regard to this number of complete turns, its mean and standard deviation are computed per each set of 20 values of it. For the gyroscope, a sampling rate of 10.2 Hz is adequate, which leads to compute a mean and standard deviation almost every 2 s. In Figure 5, the number of steering wheel complete turns, its mean and standard deviation while the app executes, are also showed.

As proposed in [36], each value of the angular speed of the steering wheel is classified as a function of the interval, out of 5, where it falls: larger than 10°, between 10° and 7.5°, between 7.5° and 5°, between 5° and 2.5°, and lower than 2.5°. From the number of samples falling within each interval, the percentage with respect to the total number of samples and the mean and standard deviation of the angular speed for each interval are shown in the lower part of Figure 5.

As proposed in [34], our app also computes the number of zero-crossing points of the angular speed and the maximum value of the absolute angular speed. Both parameters are used in [34] because an alert driver has a larger number of zero-crossing points and a lower number of the maximum value of the absolute angular speed than a drowsy driver. This is due to the fact that an alert driver continuously corrects the steering wheel position to keep the vehicle within the desirable path. Furthermore, the value of the absolute angular speed is larger in drowsy drivers as they notice a vehicle deviation from the main path with more delay than an alert driver, which requires big sudden changes in the angular speed of the steering wheel.

To implement a zero-crossing counter for the angular speed, a variable that stores the angular speed in each sampling instant is used so that it can be compared to its value in the next instant and then a sign change can be checked. Moreover, the maximum value of the absolute angular speed is stored from the beginning of the gyroscope transmission.

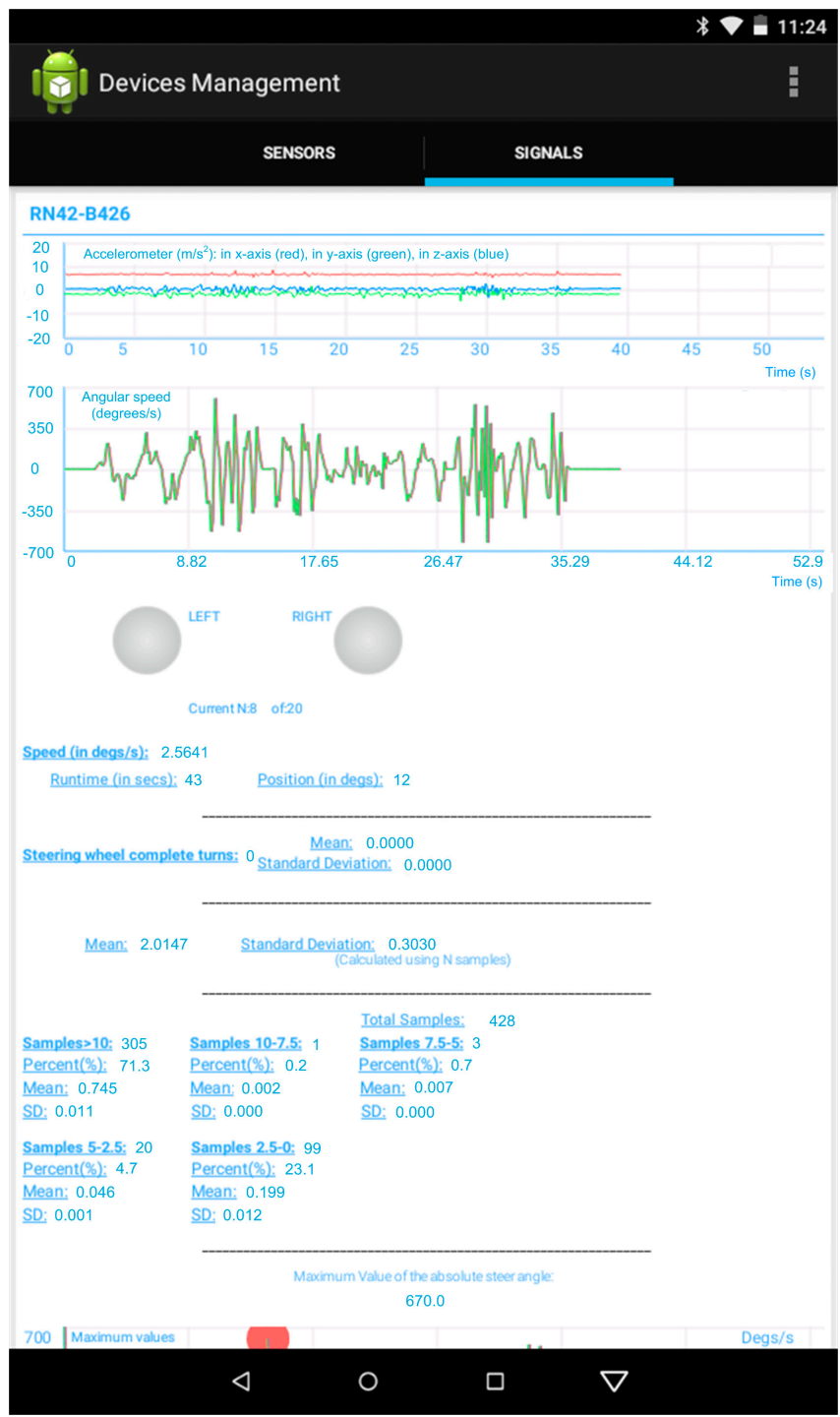

**Figure 5.** Graphs with the data extracted by the Nine Degreees of Freedom (9DOF) module and calculations made by the app.

In Figure 6, two graphs from the app representing the angular speed of the steering wheel are shown. In Figure 6a, the red circles indicate the positions where the maximum value of the absolute angular speed is updated and its present maximum value is also shown. In Figure 6b, the red circles indicate the positions where zero-crossing points are computed and a counter with the present number of zero-crossing points is also shown.

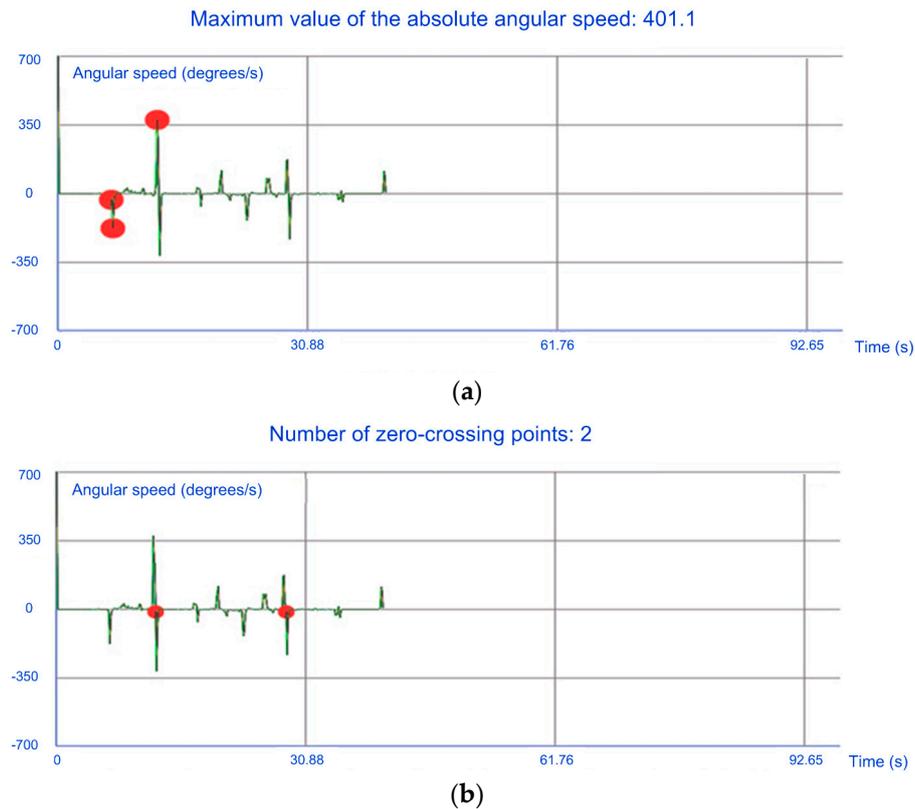

**Figure 6.** Graphs representing the angular speed of the steering wheel extracted by the app: (**a**) with the positions where the maximum value of the absolute angular speed is updated and its present maximum value; (**b**) with the positions where zero-crossing points are computed and a counter with the present number of zero-crossing points.

*2.4. Synchronization between the Android App and Our Driving Simulator*

We aimed to use the Android app for the Shimmer sensors with a driving simulator that we previously developed using the Unity game engine and C# language to analyze the driving security and efficiency. The driving simulator includes different scenarios and traffic situations of interest. The simulator stores, with a configurable frequency, a big amount of information about the vehicle in each route, such as the speed, rpm, gear, and fuel consumption. Moreover, it stores the traffic offences committed on route.

The electrodes of the ECG, EMG, and GSR sensors are placed in the proper places on the driver's body for the physiological monitoring. The 9DOF module, which includes the gyroscope, is placed on the steering wheel of the driving simulator.

We configured the Android app for the Shimmer sensors and the driving simulator to be synchronized in such a way that the beginning and end of the driving in a scenario and the beginning and end of the sensor data storage coincide. Moreover, the end of the driving in a scenario would also have to coincide with the transmission of the sensor data files to the PC where the simulator executes and in the same folder where the files with the vehicle data and traffic offences are stored. With a view to achieving this synchronization, we developed a Unity app to be installed in the Android device (bridge app). This app bridges between the Android app for the Shimmer sensors and the PC driving simulator.

The communication process is divided in two steps, first the communication between the two Unity applications (PC simulator and the bridge app), and second the communication between the PC simulator and the Android app for the Shimmer sensors.

2.4.1. Unity-Unity Communication

The two Unity applications run in different devices that need to be connected to the same Local Area Network (LAN) with access to the Internet. A standard Application Programming Interface (API), which is primarily offered by Unity to develop multiplayer games, is used for the communication of the Unity applications. The API is named Standard Unity Networking [38]. A server, which Unity offers for the net functionalities of games for free, is also used. The communication process is shown in Figure 7.

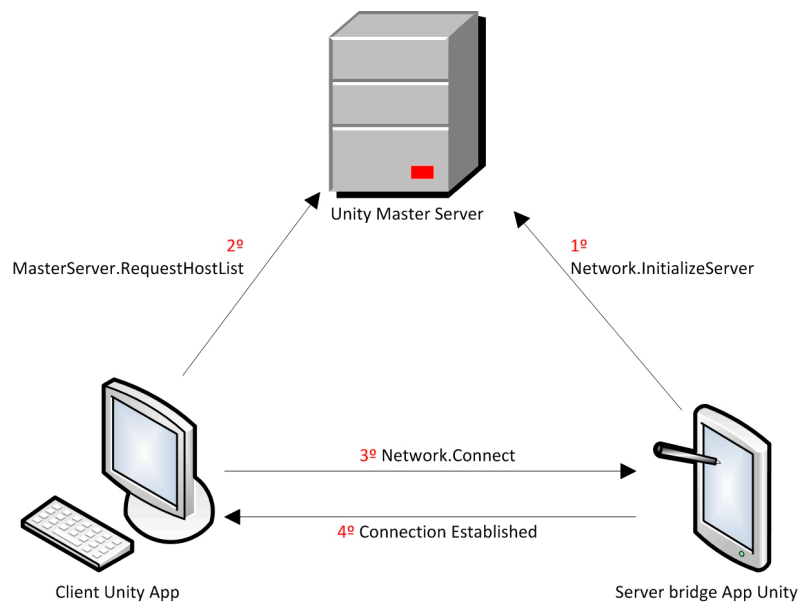

**Figure 7.** Communication between the two Unity applications.

First, in the Master Server we register our own server (*Network.InitializeServer*), which is the bridge app, indicating the name of our game. This name ("*com.ak.shimmer*") is the unique identifier. This way, the Master Server stores the correspondence between the game name and the IP address and port of our server (bridge app). Later, the client (PC Unity driving simulator) sends a request to the Master Server (*MasterServer.RequestHostList*), which returns the IP address and port of our server to the client. These data are named *HostData* in a standard way. With the *HostData*, the client establishes the communication with our server changing the client-server paradigm to a peer-to-peer connection.

2.4.2. Unity-Android Communication

To establish a communication between the PC simulator and the Android app for the Shimmer sensors, TCP sockets are used. These sockets are identified in each machine with a pair of IP addresses that identify each one of them and a pair of port numbers that identify the corresponding process in each machine. This communication is shown in Figure 8. Thanks to the bridge app and the Unity Multiplayer Networking API, the PC can know the IP address of the mobile device and vice versa, thus the socket can be established. The socket is established in the 8080 port. This port is adequate because the Android API for sockets, due to security reasons, does not allow to establish sockets in ports lower than 1024. These ports belong to more common protocols and the 8080 port is an alternative to the HTTP 80 port. It does not cause problems and firewalls do not block it. Once the connection is achieved, the bridge app can be closed or minimized and the Android app for the Shimmer sensors can be opened. This application is the server part of the socket and begins a service that listens to the previously mentioned 8080 port, waiting for the beginning of the connection from the PC. An Android server is an app component that executes in the background and can do the same actions as an activity with the only difference that the service does not provide any user interface. In the Android app for the Shimmer sensors, a button was included so that the user can select whether the sensors are going to be used in synchronization with the simulator or not. If the synchronized use is chosen, after the selection of the "Manage devices" button, a new activity opens. This activity launches a background service to establish the socket as soon as the activity is initialized. When the service is launched, a new thread is created. This thread will be in charge of listening to the given port and waiting for requests to establish the socket. With this thread, critical blocks between the main thread of the service and the application are avoided. After that, the corresponding button in the PC simulator can be clicked to send the connection request. Next, the service of the Android app for the Shimmer sensors checks if the IP address, where the request was completed, is the same IP address stored in the corresponding file. If so, the connection is accepted and the command to connect the sensors is sent to the main activity by means of a broadcast message received in a *BroadcastReceiver* component.

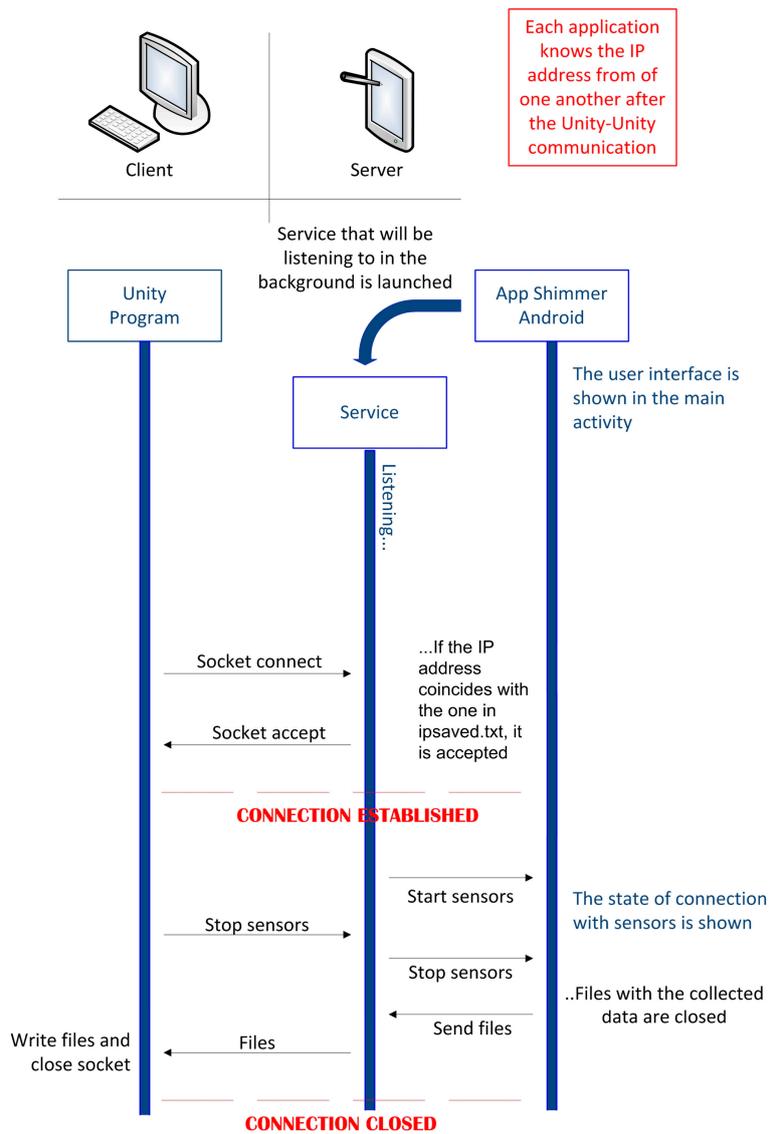

**Figure 8.** Unity-Android Communication.

From this moment forward, the main activity will keep on gathering sensor data and storing them in the corresponding files. Meanwhile, the background Android service will keep on listening to what can be received through the connection with the socket, similarly to the PC simulator. To implement this continuous listening task in Unity, the function predetermined by the Unity API (*Update*()) is used. This function is called in each frame as long as the program is executing. It is possible to stop the gathering of the sensor data by clicking the corresponding button in the PC simulator. When the service of the Android app for the Shimmer sensors receives the chain "*stopall*", a broadcast message will be sent and will be received by the main activity of the Android app for the Shimmer sensors, thanks to the implemented *BroadcastReceiver*. Thus, in the main activity, the function that stops the synchronization of the sensors will be executed and the files with the gathered data will be closed.

Once the files are closed, the command is sent to the service to send the files through the socket. A read buffer is opened to read from the files and a write buffer is opened to write the lines read from the files. After finishing the sending of the files, all the buffers and the connection with the socket are closed and the user of the Android app for the Shimmer sensors is informed that the sending of the files has finished. In the PC simulator, all data received through the socket in the Update function are gathered. Once the sending of the data is finished, the file and socket together with the connection will be closed. Moreover, although not shown in Figure 8, if a pause is selected in the driving simulator, a pause also takes place in the data gathering of the sensors until the driving simulation resumes.

## 3. Results and Discussion

### 3.1. Synchronized Working of the Sensors in the Android App

The Android application has been tested with four Android devices and with different number of sensors, sampling rates, and signals to obtain in each sensor. The first two Android devices were smartphones with basic features and running the version 2.3 of the Android operating system: Samsung Galaxy SCL (i9003) (Samsung, Seoul, Korea) and the Motorola Moto G (XT1032) (Motorola, Schaumburg, IL, USA). The former has a 1 GHz 32-bit Single Core Processor (TSMC, Hsinchu, Taiwan) and 478 MB LPDDR RAM and the latter has a 1.2 GHz 32-bit Quad Core Processor (TSMC, Hsinchu, Taiwan) and 1 GB LPDDR2 RAM. The third and fourth devices were a Wolder miTab ONE tablet running version 5.1 of the Android operating system with a 1.3 GHz Quad Core Processor and 1 GB RAM and a Samsung Galaxy J5 2017 smartphone with a Samsung Exynos 1.6 GHz 64-bit Eight Core Processor and 2 GB RAM running version 8.1 of Android. The aim of using these four devices was to study the constraints of the application as a function of the performance of devices not on the high end of the market.

The first test was the simplest and consists of using the ECG, EMG, GSR, and 9DOF sensors with a sampling rate of 10.2 Hz, 20-min duration, and synchronized transmission. The application worked correctly in the four devices without any interruption, delay, or block. As a result, it was proved that the application can communicate, represent, and store data from the 4 different sensors in a synchronized way.

The second test was with the four sensors for 20 min but with a sampling rate of 50.2 Hz for the ECG and EMG sensors and a different sampling rate of 10.2 Hz for the GSR and 9DOF sensors. While the Galaxy SCL suffered from several blocks before closing the application abruptly, the Motorola Moto G, Wolder miTab ONE, and Samsung Galaxy J5 were able to end the test.

The third test was also performed with the four sensors for 20 min but all of them working with a sampling rate of 50.2 Hz in a synchronized way. Neither the Galaxy SCL nor the Motorola Moto G worked properly. Therefore, with those two devices it is necessary to use the sampling rate of 10.2 Hz or reduce the number of Shimmer sensors. On the contrary, both the Wolder miTab ONE tablet and the Samsung Galaxy J5 smartphone worked correctly until the end of the test.

The fourth test was with the four sensors for 20 min, the ECG and EMG sensors at a sampling rate of 128 Hz and the GSR and 9DOF sensors at a sampling rate of 10.2 Hz. Similarly to the previous test, the Galaxy SCL and the Motorola Moto G did not work properly but the Wolder miTab ONE and Samsung Galaxy J5 worked properly until the end of the test.

We tested all the sensors individually with sampling rates ranging from 10.2 Hz to 128 Hz. All these tests were satisfactory as all the sensors worked properly without any interruption or block.

Lastly, we used the Samsung Galaxy J5 to make a performance testing of the app using the fourth sensors simultaneously, the ECG and EMG with a sampling rate of 128 Hz and the GSR and DOF with a sampling rate of 10.2 Hz. These rates for ECG, EMG, and GSR are suitable to monitor the physiological of a person. Regarding the 9DOF sensor placed on the steering wheel, the sampling rate of 10.2 Hz is suitable to obtain features to monitor driving performance. This final test lasted 60 min. The app starts up immediately as soon as the user taps on the app icon with no perceived delay. Once the app is running, it uses 13.5 MB of RAM memory. During the test, the battery temperature was below 31 ºC and the battery consumption after the 60 min was 5.5% of a full battery charge, being the capacity of the battery 3000 mAh. The app can run in parallel with other apps without any interference. Other apps such as WhatsApp, Gmail, and Samsung Gallery were used while our app was running in the background and it was retrieved to the foreground remaining in the same state that it was before bringing it to the background. While the app was running in the foreground, it was possible to select one of the four sensors that were sending data to observe their values and graphs in real time as described in Section 2. The selection from one sensor to another was made smoothly. Regarding the DOF sensor, whose graphs and data cannot be seen altogether in the smartphone screen, the scrolling of the screen to see all the information was made smoothly. Besides, all the graphs with sensor data can be zoomed in and out smoothly by tapping on them.

When the service of the app receives the chain "*stopall*" from the PC simulator, the files with the data from all the sensors were closed. The transmission of the files from the app to the PC simulator is made through Transmission Control Protocol (TCP) socket. This socket is connection-oriented and guarantees the reliable reception of bytes in the PC simulator in the same order that they are transmitted from the Android app.

After the 60-min test, the files with the sensor data are quite big and it took 27 s to send the files through the socket by writing in a write buffer the lines read from the files, which were then read by a read buffer of the socket. That transmission time is not problematic for the application given that the driver simulator has stopped and the sensor data are sent to the PC simulator correctly. During the test, the Android smartphone and the PC simulator were connected to a LAN of 100 Mbps data transfer rate. If the LAN did not provide access, as the files are stored in the smartphone, they could be sent later as soon as the connection is recovered.

Through the simultaneous monitoring of the ECG, EMG, and GSR of a person, the application can be applied to many different environments such as patients at their homes or in hospital or people doing some physical work or sport, or, in general terms, while doing some activity which requires a certain level of physical or psychological effort such as driving. This monitoring of ECG, EMG, and GSR of a person could be completed properly with a sampling rate of 128 Hz for ECG and EMG and of 10.2 Hz for GSR, which requires a device with features similar or higher than the Wolder mi Tab ONE tablet. Thus, if the physical or psychological conditions are below some threshold, the user or some clinician can be sent a warning to act accordingly.

We aimed to use the Android app for the Shimmer sensors to monitor drivers in our simulator. The 9DOF sensor is placed on the steering wheel and a sampling rate of 10.2 Hz is appropriate. The same sampling rate of 10.2 Hz is adequate for the GSR sensor. For the rest of sensors (ECG and EMG), the proper sampling rate is 128 Hz. With these rates, the Wolder mi Tab ONE tablet and Samsung Galaxy J5 smartphone have been proved to work properly

*3.2. Experimental Results with the Android App and Our Driving Simulator*

In this section, we show the experiments carried out with the synchronized working of the driving simulator and the Android app for the Shimmer sensors. We made two sets of experiments. In the first set, we focused on the data extracted from the gyroscope and the variables computed by the Android app from these data. In the second set, we focused on the ECG, EMG, and GSR signals.

19 participants took part in the first set of experiments. From 4 out of the 19 participants, we did not use the gyroscope in their scenario routes so only data from the simulator were stored. From the remaining 15 participants, both the data from the gyroscope using the Android app and from the simulator were stored. The mean age of the participants was 26.68 years old with a standard deviation of 6.77 years old. The mean length of time for the participants having a driving license was 7.79 years with a standard deviation of 6.09 years. Only one of the 19 participants did not have a driving license although he or she was preparing to get it. The participants were asked to rate themselves with a number between 1 and 10 according to their experience in computer games and according to their experience in computer racing games. The mean value of the first ranked experience was 6.53 with a standard deviation of 3.43 and the mean value of the second ranked experience was 4.53 with a standard deviation of 3.23.

Each participant drove in four different scenarios of the simulator twice, the first time using automatic gear shift and the second time using manual gear shift. The four scenarios were developed as varied as possible and covering the different range of roads, traffic, and situations that have to be faced in real driving situations and require different driving skills. The first scenario was mainly urban with an interurban section consisting in a winding road with one lane per direction. The second scenario was interurban in a ring road in the surroundings of a big city and with a variable number of lanes per direction. The third scenario was urban in a through road on the outskirts of a medium-sized city. The fourth scenario was a monotonous route in a highway. The approximate time to complete the first, third, and fourth scenarios was 10 min and the approximate time to complete the second scenario was 20 min. As the participants completed each scenario twice, the approximate mean time to finish with all the scenarios was 100 min for each participant.

Before driving in the simulator, the participants had to make two questionnaires to have a subjective measure of their sleepiness at that very moment. These two questionnaires were the Karolinska Sleepiness Scale (KSS), and the Stanford Sleepiness Scale (SSS). Moreover, the participants made the Epword Sleepiness Scale (ESS), which evaluates the likelihood of falling asleep not at the time of making the questionnaire but in different real situations such as watching TV or being sat in a public place. Those questionnaires have been adopted in numerous research studies related to the clinical assessment of fitness to drive in sleepy individuals using a driving simulator [39]. We aimed to relate the results obtained in these questionnaires with the driving performance computed with different variables.

The steering wheel adopted in the simulator was the Logitech G27 that also includes the gear lever and the clutch, brake, and accelerator pedals, making it possible a realistic driving experience in the simulator. The simulator scene is projected onto a big screen of dimensions 260 × 195 cm as shown in Figure 9.

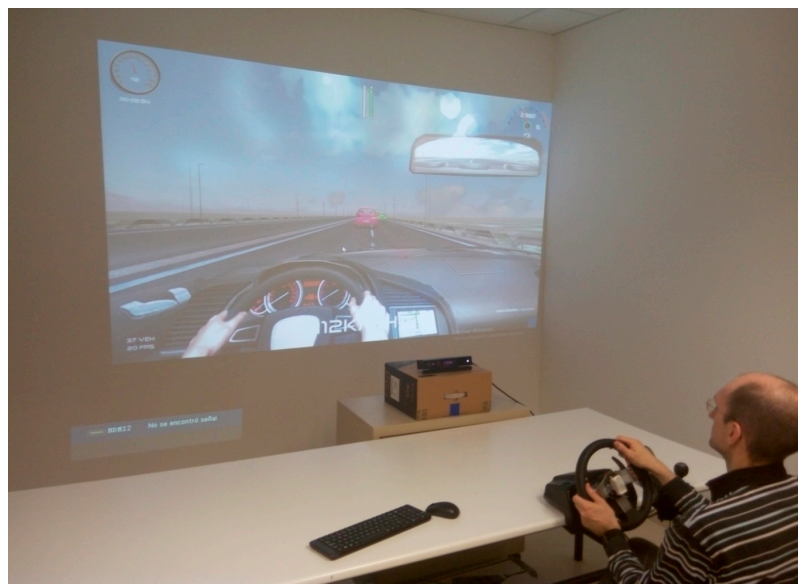

**Figure 9.** Driving simulator.

The simulator was configured to store the following vehicle data every second: instantaneous fuel consumption, speed, rpm, and position in the scenario. Moreover, the time and position in the scenario of all the committed traffic offences are stored. The implemented traffic offences were: collision with a pedestrian, collision with a vehicle, collision with a motorcycle, collision with a cyclist, collision with a still object, leaving the road, over-speed, under-speed, leaving a roundabout incorrectly, driving in a roundabout incorrectly, leaving a junction incorrectly, failing to comply with a stop sign, failing to comply with a yield sign, failing to comply with a traffic light signal, not respecting the safety distance, crossing a solid line in a road, not using the turn light in a turn, and not using the turn light while overtaking a car.

In the first set of experiments, the Android app for the Shimmer sensors stored the angular speed obtained from the gyroscope placed on the steering wheel. The angular speed was obtained from the gyroscope with a sampling rate of 10.2 Hz. From this speed, many variables were computed by the Android app as explained in Section 2.3. For the data analysis, we used the software MATLAB (Natick, MA, USA), IBM SPSS (Armonk, NY, USA), and Excel (Redmond, WA, USA).

We first analyzed the influence of the type of gear shift on the speed and the variables extracted from the angular speed of the gyroscope. We aimed to test whether the gear shift has some influence on the driving style as, on the one hand, not all the participants have driven a car with the automatic gear shift in their driving history and, on the other hand, none of them is used to drive in a simulator.

We split the analysis into two parts, the first part grouping together the data from the two urban scenarios and the second part grouping together the data from the two interurban scenarios. Figure 10; Figure 11 show the mean speed of the 19 participants in the urban scenarios and in the interurban scenarios, respectively.

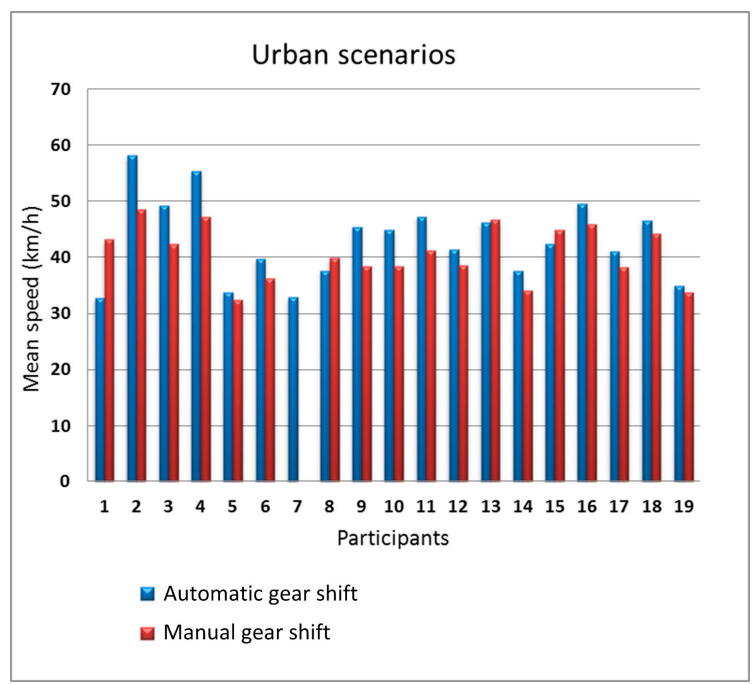

**Figure 10.** Mean speed as a function of the gear shift type in the urban scenarios.

In the urban scenarios, there is some influence of the type of gear shift on the speed. 14 people drove faster with the automatic gear shift (participants #2, #3, #4, #9, #10, and #11 with more sensitive differences). Only four participants drove faster with manual gear shift but with a small difference with respect to automatic gear shift. Furthermore, for the interurban scenarios, the difference between the speed with manual gear shift and automatic gear shift is small, being the number of participants that drove faster with automatic gear shift equal to the number of participants that drove faster with manual gear shift. In the urban scenarios, gear shifts are more frequent so, logically, the driving differences using the two types of gear shift is larger in those scenarios.

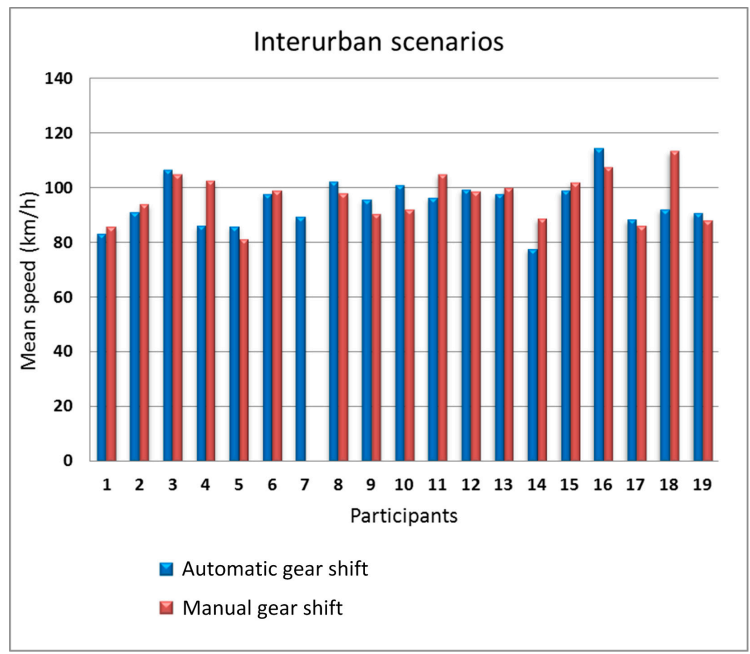

**Figure 11.** Mean speed as a function of the gear shift type in the interurban scenarios.

We analyzed the relation between the gear shift type and the mean and the standard deviation of the angular speed obtained from the gyroscope for the urban and interurban scenarios. Figures 12 and 13 show the mean angular speed of the steering wheel in the urban and interurban scenarios, respectively, of the 15 participants that used the gyroscope in the simulations.

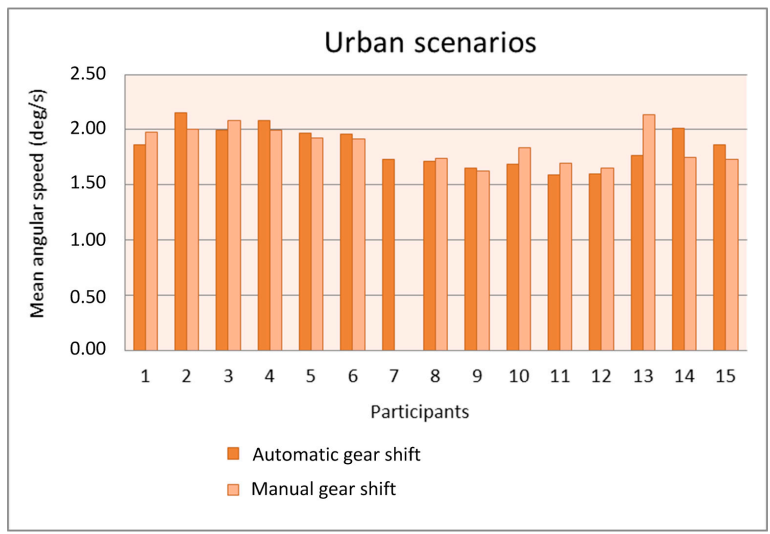

**Figure 12.** Mean angular speed as a function of the gear shift type in the urban scenarios.

With the exception of participant 13, the mean is quite similar with the automatic gear shift and the manual gear shift. The number of participants with a mean larger with the automatic gear shift is similar to the number of participants with a mean larger with the manual gear shift. The mean values in the urban scenarios are similar to the mean values in the interurban scenarios and, only for some participants, these values are larger in the interurban scenarios. The use of the manual gear shift did make participant #13 increase the mean angular speed because that gear shift distracted him or her from driving the car and he or she had to correct the trajectory with more motions of the steering wheel. Anyway, this lack of adaptation to the manual gear shift in the simulator was only present in that participant.

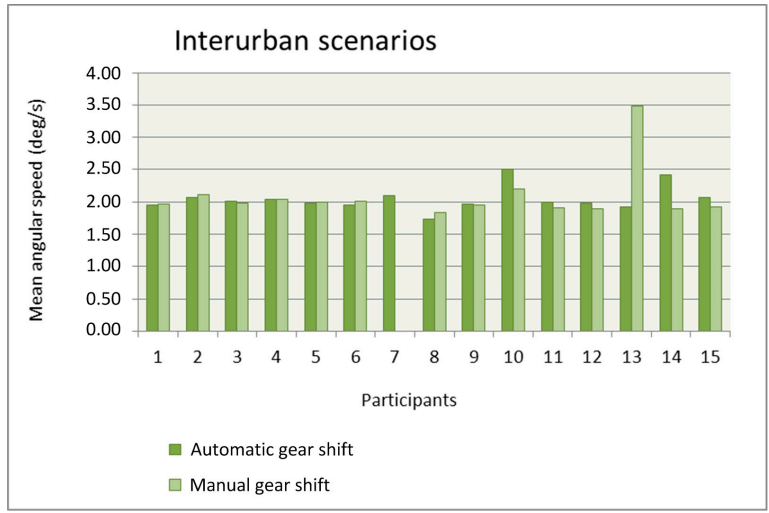

**Figure 13.** Mean angular speed as a function of the gear shift type in the interurban scenarios.

Figures 14 and 15 show the standard deviation of the angular speed of the steering wheel in the urban and interurban scenarios, respectively. In general, the value of the standard deviation is not influenced by the adopted gear shift. Whereas in the urban scenarios, eight participants have a larger standard deviation with automatic gear shift and six participants have a larger one with manual gear shift, in the interurban scenarios six participants have a larger standard deviation with automatic gear shift and seven participants have a larger one with manual gear shift. The value of the standard deviation is larger in the urban scenarios as the drivers have to make sharper turns in urban scenarios than in interurban scenarios.

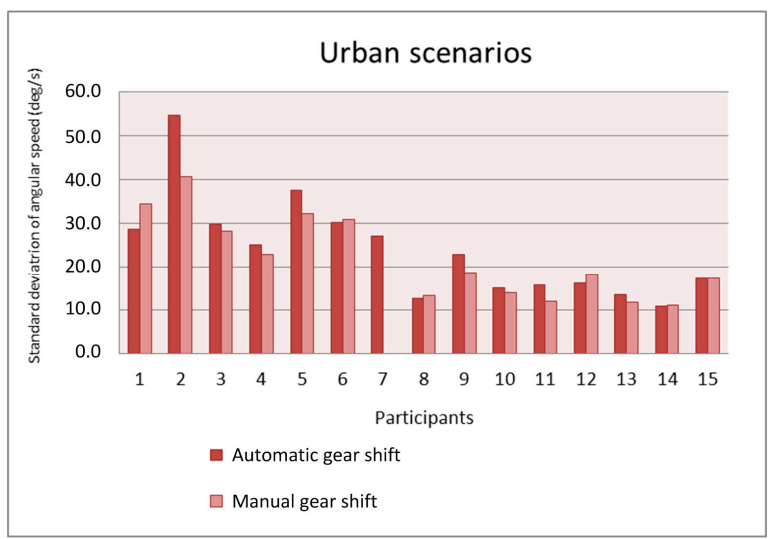

**Figure 14.** Standard deviation of the angular speed as a function of the gear shift type in the urban scenarios.

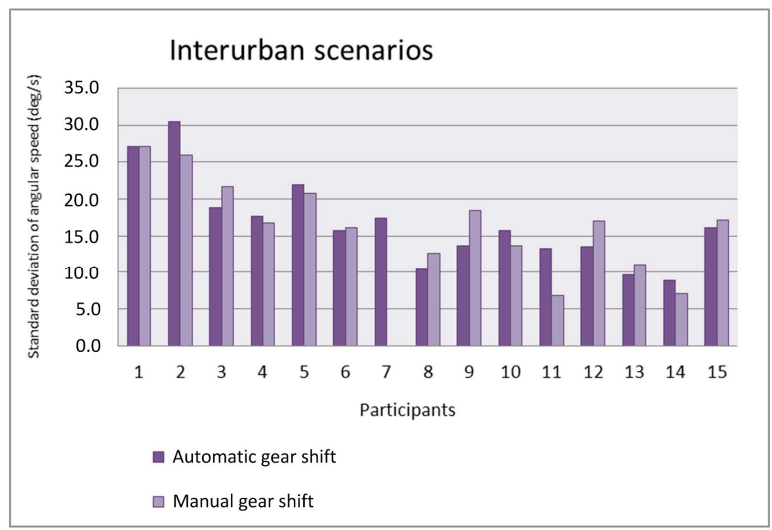

**Figure 15.** Standard deviation of the angular speed as a function of the gear shift type in the interurban scenarios.

We also analyzed the relation between the mean and the standard deviation of the angular speed of the steering wheel in the urban and interurban scenarios and the traffic offences committed. The time spent to finish a route varied from one participant to another as they drove at different speeds and some participants chose the wrong exit in some junctions and drive more distance than others. Because of this, we analyzed the number of traffic offences per second in each scenario instead of the total number of traffic offences. In Table 1, the participants are ordered as a function of the decreasing number of traffic offences committed per minute in the urban scenarios and with the mean and standard deviation of the angular speed. In Table 2, the participants are ordered with the same criteria but in the interurban scenarios.

**Table 1.** Participants ordered as a function of the number of traffic offences per second in urban scenarios.

| | Urban Scenarios | | |
|---|---|---|---|
| Participants | Traffic Offences/s | Mean Angular Speed | Standard Deviation of the Angular Speed |
| 7 | 0.142 | 1.731 | 27.048 |
| 14 | 0.109 | 1.880 | 11.016 |
| 1 | 0.092 | 1.921 | 31.524 |
| 2 | 0.085 | 2.080 | 47.617 |
| 3 | 0.071 | 2.038 | 28.986 |
| 10 | 0.063 | 1.760 | 14.505 |
| 4 | 0.062 | 2.036 | 23.957 |
| 8 | 0.055 | 1.724 | 12.906 |
| 13 | 0.046 | 1.952 | 12.706 |
| 5 | 0.045 | 1.945 | 34.903 |
| 12 | 0.032 | 1.622 | 17.254 |
| 15 | 0.030 | 1.798 | 17.292 |
| 11 | 0.029 | 1.644 | 13.892 |
| 6 | 0.027 | 1.940 | 30.543 |
| 9 | 0.026 | 1.640 | 20.727 |

**Table 2.** Participants ordered as a function of the number of traffic offences per second in interurban scenarios.

| | Interurban Scenarios | | |
|---|---|---|---|
| Participants | Traffic Offences/s | Mean Angular Speed | Standard Deviation of the Angular Speed |
| 2 | 0.1023 | 2.085 | 28.251 |
| 4 | 0.0727 | 2.043 | 17.206 |
| 3 | 0.0683 | 2.000 | 20.252 |
| 6 | 0.0629 | 1.981 | 15.927 |
| 1 | 0.0570 | 1.966 | 27.092 |
| 12 | 0.0508 | 1.935 | 15.324 |
| 13 | 0.0464 | 2.701 | 10.413 |
| 9 | 0.0438 | 1.961 | 16.108 |
| 11 | 0.0428 | 1.952 | 10.051 |
| 7 | 0.0388 | 2.100 | 17.414 |
| 10 | 0.0360 | 2.354 | 14.759 |
| 15 | 0.0355 | 1.997 | 16.607 |
| 14 | 0.0326 | 2.157 | 7.986 |
| 8 | 0.0247 | 1.789 | 11.534 |
| 5 | 0.0079 | 1.986 | 21.355 |

It can be observed that there is no significant relation between the number of traffic offences per second and the mean and standard deviation of the angular speed for the urban and interurban scenarios.

Besides, we studied the percentage of mean values of the angular speed that lied within one of the five intervals considered: larger than 10°, between 10° and 7.5°, between 7.5° and 5°, between 5° and 2.5°, and lower than 2.5°. Studying these percentages for the users with larger and lower means, we did not find any significant data that could characterize them or relate them to the total number of traffic offences. Then, we analyzed the mean number of zero-crossing points per second in both the urban and interurban scenarios looking for some relation between them, the traffic offences, and the mean speed. We did not observe relations between these variables but what we did observe was that for the most participants, the mean number of zero-crossing points per second was larger for the urban scenarios than for the interurban scenarios, as shown in Figure 16. In the interurban scenarios, the position of the steering wheel has to be modified less frequently than in the urban scenarios.

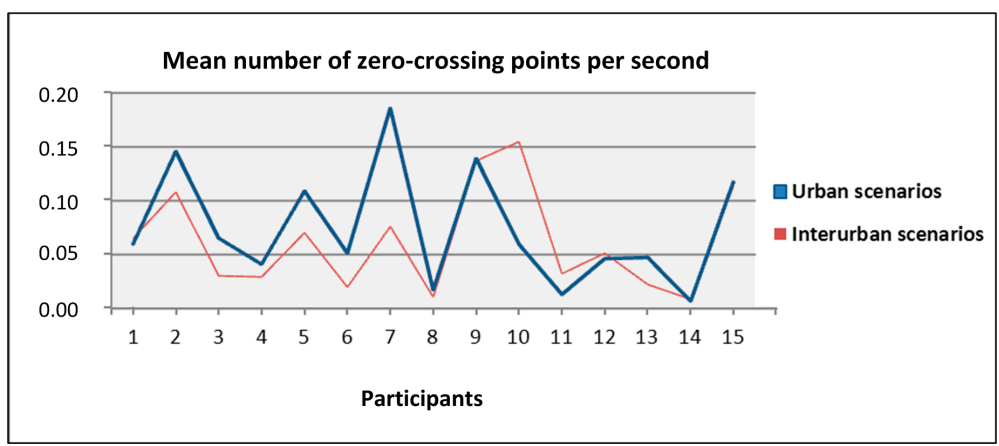

**Figure 16.** Mean number of zero-crossing points per second for urban and interurban scenarios per each participant.

We conducted a detailed statistical study for each type of traffic offence in the urban and interurban scenarios looking for its correlation with the values of the mean and standard deviation of the angular speed obtained from the gyroscope and the speed, rpm, and mean fuel consumption in each simulation. In this study, first we used the Shapiro-Wilk test [40]. This test is usually employed to verify if a set of data follows a normal or Gaussian distribution. If at least one of the two variables that one wants to correlate follows a normal distribution, the Pearson correlation coefficient can be used [41]. For the Shapiro-Wilk test, the null hypothesis, which is aimed to be rejected, is that a sample $x_1, x_2, \ldots, x_n$ comes from a normally distributed population. Once the test has been completed, if the *p*-value is larger than $\alpha$ (significance level is set to 0.05), the null hypothesis is not rejected concluding that the data follow a normal distribution. Applying the Shapiro-Wilk test to the variables mean and standard deviation of the angular speed, speed, rpm and mean fuel consumption, we obtained *p*-values of 0.23, 0.18, 0.45, 0.21, and 0.31, respectively. In all the cases, the values were larger than $\alpha$ (0.05) so that all the variables followed a normal distribution and we could correlate them with different types of traffic offences using the Pearson correlation coefficient.

The Pearson correlation coefficient is a measure of the linear correlation between two quantitative random variables and is independent of their measuring range. It is calculated as follows:

$$\rho_{XY} = \frac{\sigma_{XY}}{\sigma_X \sigma_Y}, \qquad (2)$$

where $\sigma_{XY}$ is the covariance between *X* and *Y*, $\sigma_X$ is the standard deviation of *X*, and $\sigma_Y$ is the standard deviation of *Y*. Its values lie within the interval $[-1, 1]$, being 1 if there is a perfect positive linear relation between the variables, 0 if there is no linear correlation between them, and $-1$ if there is a perfect negative linear relation between them.

In our statistical study, we analyzed the data from driving with both automatic and manual gear shifts together. On the contrary, we have performed one independent analysis for the urban scenario and another one for the interurban scenarios. We took that approach as there are traffic offences that can only be present in the urban scenarios such as leaving a roundabout incorrectly and driving in a roundabout incorrectly as there are no roundabouts in the interurban scenarios in our simulator.

After calculating all the Pearson correlation coefficients between each one of the five variables (two obtained from the Android app for the Shimmer sensors and three obtained from vehicle performance in the simulator) and each type of traffic offence differentiating urban and interurban scenarios, we tested their statistical significance calculating the *p*-value. The *p*-value in this case is the probability of having obtained such values of the Pearson correlation coefficients supposing that the null hypothesis is true, that is, there is no linear correlation between the pair of variables. If the *p*-value is lower than $\alpha$ (typically 0.05), the null hypothesis is rejected and the statistical significance of the correlation value is verified.

Tables 3 and 4 show only the correlation coefficients that have led to *p*-values lower than 0.05 for the urban and interurban scenarios, respectively. There are more significant correlations in the urban than in the interurban driving. The only valid correlation with a variable extracted from the gyroscope was the standard deviation of the angular speed in urban scenarios with the traffic offence of leaving the road. Hard turns of the steering wheel, which increase the value of the standard deviation of the angular speed and are typical of bad driving, are related to leaving the road in an urban environment. The other three verified correlations have included the traffic offence of not using the turn light in a turn, two cases in an urban scenario (for mean speed and mean rpm) and one case in an interurban one (for mean fuel consumption). In the urban scenarios, as the speed or the rpm increase, the probability of turning without using the corresponding traffic light also increases. Although it may seem that this traffic offence is not serious, it can cause dangerous situations or even traffic accidents due to the lack of information that the rest of the vehicles have. In the interurban scenarios, the variable correlated with the traffic offence of not using the turn light in a turn was the mean fuel consumption. It may seem that several variables, other than the mean fuel consumption, are more related to this traffic offence. However, the mean fuel consumption is representative of the driving style in interurban scenarios and was interestingly correlated with the traffic offence of not using the turn light in a turn in our experiments.

**Table 3.** Pearson correlation coefficients with *p*-values lower than 0.05 for the urban scenarios.

|  | Urban Scenarios | |
| --- | --- | --- |
|  | **Traffic Offence** | **Pearson Correlation Coefficient** |
| Standard deviation of the angular speed | Leaving the road | 0.46 |
| Mean speed | Not using the turn light in a turn | 0.54 |
| Mean rpm | Not using the turn light in a turn | 0.61 |

**Table 4.** Pearson correlation coefficients with *p*-value lower than 0.05 for the interurban scenarios.

|  | Interurban Scenarios | |
| --- | --- | --- |
|  | **Traffic Offence** | **Pearson Correlation Coefficient** |
| Mean fuel consumption | Not using the turn light in a turn | 0.45 |

Following the study, we analyzed the relation between the number of traffic offences per second in the simulations and the length of time having a driving license, experience in computer games, and experience in computer racing games of the participants in the experiments. First, we verified that the number of traffic offences per second has a normal distribution with the Shapiro-Wilk test. Then, we obtained the Pearson correlation coefficient between the traffic offences per second and the three mentioned variables related to the driver experience. The *p*-value of these correlations was larger than 0.05 in the three cases so no significant correlation was found. A significant correlation between the number of traffic offences and the experience in computer games or computer car games would have meant a relation between this experience and the performance in the simulator. As the simulator has been developed to be as similar as possible to real driving and not to computer games, this lack of correlation can be considered a positive aspect of the simulator.

Next, we analyzed the relation between the subjective assessment of participants' sleepiness through KSS, SSS, and ESS questionnaires and a measure of low vigilance in the simulations. We used a measure similar to the one proposed in [34]. They related a lower number of zero-crossing points of the steering wheel to a less vigilant driver that does not make continuous direction corrections. They also related maximum values of the absolute angular angle with a drowsy driver that has to do particular hard turns.

From these measures, we considered a possible driving period of low attention if, in this period or the previous one, there was no zero-crossing point and a maximum value of the absolute steer angle has been obtained. We divided the simulations in periods of 5 s so that each period could be classified as having or not having possible low attention driving. With a sampling rate of 10.2 Hz for the gyroscope, there are 51 samples of zero-crossing and maximum value of the absolute angular speed in each period. We obtained the Pearson correlation coefficient between the number of low attention periods and the results obtained in KSS, SSS, and ESS. We also obtained the Pearson correlation coefficient between the number of low attention periods and the age, length of time having a driving license, and experience in computer racing games. The *p*-value of these correlations was larger than 0.05 for all except for the length of time having a driving license, whose Pearson correlation coefficient was 0.5594. The more experienced drivers had more low attention periods, which should be avoided in safe driving, in the simulations.

Finally, we studied the correlation between the number of traffic offences in all the scenarios and the results in the KSS, SSS, and ESS. The only Pearson correlation coefficient with a *p*-value lower than 0.05 was with SSS, whose value was 0.5375. SSS is a scale of seven statements related to an increasing level of sleepiness. We found this valuable relation between a higher subjective assessment of the level of sleepiness and a larger global number of traffic offences although we did not find this relation with the number of low attention periods computed as in [34].

In the second set of experiments, in which we focused on the ECG, EMG, and GSR signals, 6 participants took part. The mean age of the participants was 32.83 years old with a standard deviation of 16.82 years old. The mean length of time for the participants having a driving license was 20.83 years with a standard deviation of 14.85 years. They were required to drive in the scenarios twice, one time when they were rested and another time when they were tired. They were free to choose the time and day representative of these two physical states to drive in the scenarios.

Figure 17 shows the mean value of HRV obtained from the ECG sensor when the participants drove in the scenarios in rested state and distinguishing between the urban and interurban scenarios. Figure 18 shows the similar information but when the participants drove in the scenarios in tired state. Most of the participants had bigger differences in the mean HRV, as a function of the scenarios (urban or interurban), in rested state than in tired state. For some participants, the mean HRV is bigger in the urban scenarios than in the interurban scenarios and for others is the other way round. That way, as there is a relation between the HRV and the physiological state, it can be asserted that in rested state most of the participants somehow adapted their physiological states depending on the requirements of each scenario while in the tired state the different scenarios did not make participants modify their physiological states significantly. There are big differences between participants so that an analysis to determine if the physiological or psychological state is suitable to drive would have to be individually specified from the learning of the relation between the HRV and the different scenarios and traffic states.

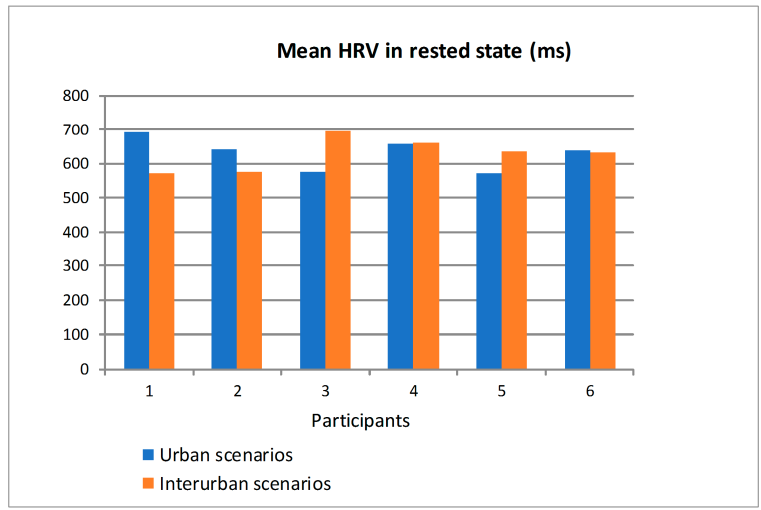

**Figure 17.** Mean HRV of the participants in rested state driving in urban and interurban scenarios.

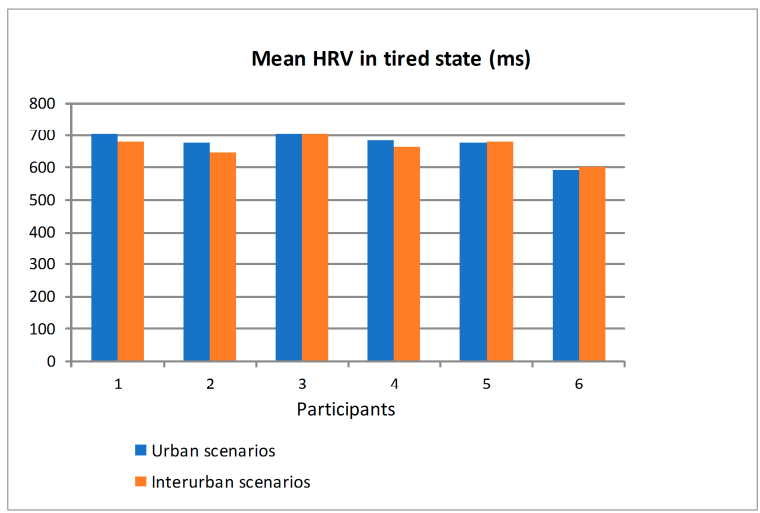

**Figure 18.** Mean Heart Rate Variability (HRV) of the participants in tired state driving in urban and interurban scenarios.

The electrodes of the EMG sensor were placed on the trapezius muscle. We chose that muscle as it is especially used while moving the steering wheel. Figure 19 shows the mean value of EMG in mV (millivolts) for the six participants distinguishing between driving in rested state and tired state. For all the participants, the mean EMG was bigger in tired state so that the trapezius sends electrical impulses of bigger amplitudes as someone gets tired. The differences in the mean EMG in both states were very irregular depending on the participant so that a personalized threshold should be established to detect the transition from a physiological state suitable to drive to a tired state that would eventually lead to unsafe driving based on this parameter.

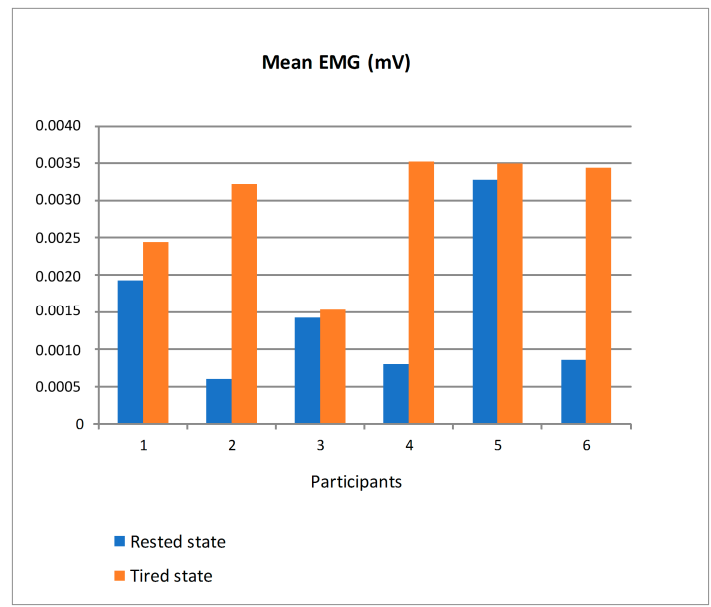

**Figure 19.** Mean EMG of the participants in rested and tired state.

Figure 20 shows the mean value of GSR in kOhm for the six participants distinguishing between driving in rested state and in tired state. There are clear differences depending on the participants but two tendencies can be observed. On the one hand, participants #2 and #6 had a quite similar mean GSR in both states. On the other hand, the remaining participants had a clearly bigger mean GSR in tired state. For the latter, driving in tired state made them change the emotional state and so the skin resistance captured by the GSR sensor. Similarly to the ECG and EMG, a personalized analysis would need to be applied to establish a threshold in the GSR value to detect the transition from safe to unsafe driving based on this parameter.

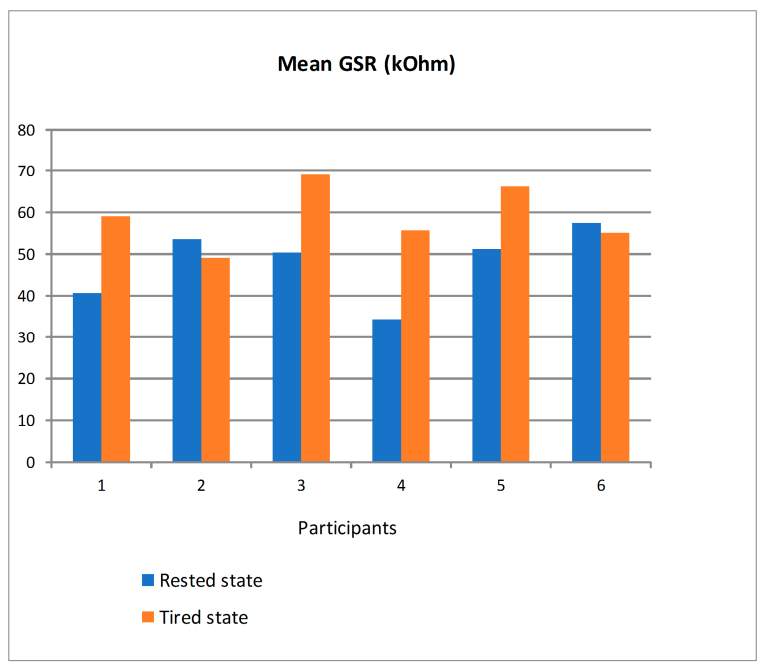

**Figure 20.** Mean GSR of the participants in rested and tired state.

## 4. Conclusions

In this paper, we present an Android mobile application for the control of the physiological sensors from the Shimmer platform and the processing and monitoring of their signals. The application is compatible with version 2.3 and above of the Android operating system.

The connection with the sensors is transparent to the user. The application receives the sensor signals, processes and represents them graphically and stores them in files for later analysis. It can use four sensors simultaneously in a low-cost Android device, the ECG and EMG sensors at a sampling rate of 128 Hz and the GSR and gyroscope sensors at a sampling rate of 10.2 Hz. These rates are sufficient for proper health monitoring so the application can be applied to many interesting environments.

The Android app is synchronized with a Unity-based driving simulator and processes a large number of features from the ECG, EMG, GSR, and gyroscope signals. Thus, the app can be used in driving research studies in a simulator. From the analysis of the experiments carried out by 25 people using the driving simulator and the Android app, different findings were achieved. Regarding the ECG, EMG, and GSR signals, significant differences were found depending on the driver's physical state (rested or tired). Personalized analyses need to be accomplished with these signals so that they can be used to detect if the driver's physiological state can lead to safe driving. Regarding the gyroscope signal, some findings were expected as the similarity between the mean angular speed of the steering wheel in urban and interurban scenarios and the larger value of the standard deviation of the angular speed of the steering wheel in urban than in interurban scenarios. Other findings were unexpected as the correlation between the traffic offence of not using the turn light in a turn (instead of others) and vehicle data such as mean speed in urban scenarios and mean fuel consumption in interurban scenarios. Overall, in urban scenarios the driving style is more associated with some traffic offences than in interurban scenarios. There was no relation between the experience in computer racing games and the number of traffic offences, which somehow validates the use of the simulator for drivers with different backgrounds. There was also a significant correlation between the number of traffic offences and the results in the SSS questionnaire, which supports the use of such questionnaire in research studies related with driving safety. Most of the analyzed correlations between variables have not been significant as there is a great range of differences between habits and performance of drivers and detailed individual analyses could be insightful.

The Android app can be used in research studies in real routes given than the vehicle data are obtained from its On Board Diagnostics (OBD) system. More experiments are needed with both the driving simulator and in real routes to analyze the physical and emotional state of the driver from the ECG, EMG, and GSR sensors, the gyroscope features and the vehicle data. From this analysis, both joining data from all the drivers and specifying it for each driver, the Android app can be extended so that a real-time processing of the signals would be able to alert the user as soon as some data imply an abnormal physical or emotional state. For instance, this functionality can alert a driver when the level of attention and drowsiness are not suitable for a safe driving. The potential benefits of mobile applications with this functionality in our lives are great and numerous.

**Author Contributions:** Conceptualization, D.G.-O. and F.J.D.-P.; Methodology, D.G.-O. and F.J.D.-P.; Software, D.G.-O., M.M.-Z. and M.A.-R.; Experimental results, D.G.-O., M.M.-Z. and M.A.-R., Supervision, D.G.-O.; Writing—original draft, D.G.-O., F.J.D., M.M.-Z. and M.A.-R.

**Funding:** This work was funded by the National Department of Traffic (DGT, Dirección General de Tráfico) of the Ministry of the Interior (Spain) under research project SPIP2017-02257.

**Conflicts of Interest:** The authors declare no conflict of interest.